# Visualization of Chiral Electronic Structure and Anomalous Optical Response in a Material with Chiral Charge Density Waves

H. F. Yang[1*], K. Y. He[2*], J. Koo[3*], S. W. Shen[1], S. H. Zhang[1], G. Liu[2], Y. Z. Liu[3], C. Chen[1,4], A. J. Liang[1,5], K. Huang[1], M. X. Wang[1,5], J. J. Gao[6], X. Luo[6], L. X. Yang[7], J. P. Liu[1,5], Y. P. Sun[6,8,9], S. C. Yan[1,5], B. H. Yan[3†], Y. L. Chen[1,4,5,7†], X. Xi[2,9†], and Z. K. Liu[1,5†]

[1]*School of Physical Science and Technology, ShanghaiTech University, Shanghai 201210, P. R. China*
[2]*National Laboratory of Solid State Microstructures and Department of Physics, Nanjing University, Nanjing 210093, P. R. China*
[3]*Department of Condensed Matter Physics, Weizmann Institute of Science, Rehovot 76100, Israel*
[4]*Department of Physics, University of Oxford, Oxford, OX1 3PU, UK*
[5]*ShanghaiTech Laboratory for Topological Physics, Shanghai 201210, P. R. China*
[6]*Key Laboratory of Materials Physics, Institute of Solid State Physics, Chinese Academy of Sciences, HFIPS, Hefei 230031, P. R. China*
[7]*State Key Laboratory of Low Dimensional Quantum Physics and Department of Physics, Tsinghua University, Beijing 100084, P. R. China*
[8]*High Magnetic Field Laboratory, Chinese Academy of Sciences, HFIPS, Hefei, 230031, P. R. China*
[9]*Collaborative Innovation Center of Advanced Microstructures, Nanjing University, Nanjing 210093, P. R. China*

*Contributed equally.
†Email: binghai.yan@weizmann.ac.il, yulin.chen@physics.ox.ac.uk, xxi@nju.edu.cn, liuzhk@shanghaitech.edu.cn

**Chiral materials have attracted significant research interests as they exhibit intriguing physical properties, such as chiral optical response, spin-momentum locking and chiral induced spin selectivity. Recently, layered transition metal dichalcogenide 1$T$-TaS$_2$ has been found to host a chiral charge density wave (CDW) order. Nevertheless, the physical consequences of the chiral order, for example, in electronic structures and the optical properties, are yet to be explored. Here, we report the spectroscopic visualization of an emergent chiral electronic band structure in the CDW phase, characterized by windmill-shape Fermi surfaces. We uncover a**

**remarkable chirality-dependent circularly polarized Raman response due to the salient chiral symmetry of CDW, although the ordinary circular dichroism vanishes. Chiral Fermi surfaces and anomalous Raman responses coincide with the CDW transition, proving their lattice origin. Our work paves a path to manipulate the chiral electronic and optical properties in two-dimensional materials and explore applications in polarization optics and spintronics.**

Chiral objects are ubiquitous in nature, from gigantic galaxy to microscopic DNA structure[1-3]. They host fundamental physics (such as parity violation of sub-atomic weak interaction) and are essential for life (homo-chirality of natural bio-molecules)[1-3]. In condensed-matter systems, materials with chiral electronic states exhibit intriguing optical, magnetic, and transport properties and have attracted great research interests[1,2,4-7]. In principle, crystals with a chiral lattice structure naturally exhibit chiral electronic band structures. One example is the chiral topological semimetals, whose chiral crystal symmetry enforces chiral surface Fermi arcs that are expected to display numerous exotic physical phenomena such as circular dichroism, unusual photogalvanic effect, and nonreciprocal transport[4,8-11].

The chiral electronic structure can emerge alternatively in materials with achiral crystal structure. Due to the limited material choice, such emergent chiral electronic structure and its unique physical properties has not been investigated. The layered transition metal dichalcogenide $1T$-TaS$_2$ provides such an opportunity. It hosts an achiral crystal structure with both inversion and mirror symmetries preserved in its high-temperature normal state (Fig. 1a). Upon cooling down, $1T$-TaS$_2$ transits to an incommensurate CDW (I-CDW) phase at around 550 K, nearly-commensurate CDW (NC-CDW) phase at around 350 K, and finally commensurate CDW (C-CDW) phase at 180 K[12-14]. The C-CDW phase features commensurate lock-in of star-of-David clusters under $\sqrt{13} \times \sqrt{13}$ reconstruction (Fig. 1b), in which the reconstructed superlattice rotates by 13.9° with respect to the pristine lattice (Figs. 2c and 2d), thus breaking the mirror symmetry while preserving the inversion symmetry, which is addressed as the chiral CDW phase. The chiral CDW state is investigated for intriguing phenomena, such as cluster Mott insulator[15,16], quantum spin liquid[17-19], superconductivity under pressure[20], carrier doping[21-23], and isovalent substitution[24-26] (Fig. 1c). Because of the existence of inversion symmetry, such a chiral CDW exhibits no circular dichroism which can probe the conventional chirality. However, the direct physical impacts, e.g., the chiral electronic structure and the consequent optical properties, of the chiral order have not been addressed[25,27-38].

In this work, by performing angle-resolved photoemission spectroscopy (ARPES) measurements on $1T$-TaS$_2$ with high quality, we discover a chiral electronic structure emergent in the chiral CDW phase of $1T$-TaS$_2$ (Fig. 1c). The chiral electronic structure appears as windmill shape Fermi contours with two distinct handedness on different samples/domains. Remarkably, although the chiral electronic structure shows no circular dichroism in optical reflectivity, we observe anomalous, chirality-dependent circularly polarized Raman response, induced by the CDW-induced symmetry breaking. Furthermore, across the CDW phase transition at 350 K, we observe a simultaneous chiral transition in the electronic structure and Raman response, thus proving their common lattice origin as the chiral CDW order. Our work reveals a mechanism to generate chiral electronic structures with unique anomalous Raman responses. The chiral electronic structures in 1$T$-TaS$_2$ host application potentials on polarization optics and spintronics[39-41], nano-electronics based on mirror domain walls[42], CDW-based memory[42], stereo-chemistry and pharmacology[43,44], etc.

Fig. 2a plots the measured band structure of 1$T$-TaS$_2$ in the C-CDW phase. From the Fermi energy to high-binding-energy, the constant energy contours (CECs) clearly demonstrate counter-clockwise windmill shape features, as shown by three representative CECs at different binding energies (Fig. 2a(i-iii)). These windmill shape CECs are nicely reproduced by our *ab-initio* calculations (Fig. 2a(iv-vi)). We rule out the origin of the windmill shape being the photoemission matrix element effect by systematic polarization and photon-energy dependent ARPES measurements (Supplementary Note S3 and Fig. S1). As the windmill shape features break the mirror symmetry (cannot be overlapped with their mirror images) (Fig. 1a), we could describe them as chiral electronic structures[1,44].

Interestingly, the other type of chiral electronic structure rotating clockwise (Fig. 2b) could be obtained from another sample (and also from different regions on the same sample, see Supplementary Fig. S2). The three representative CECs (Fig. 2b(i-iii)) nicely match the mirror images of Fig. 2a(i-iii), suggesting they are of the same physical origin only with opposite chirality.

We attribute the origin of the chiral electronic structure to the emergent chiral C-CDW state, since 1$T$-TaS$_2$ crystallizes in the achiral space group of P-3m1 (No.164) with both mirror and inversion symmetries (Fig. 1a). The commensurate lock-in of star-of-David clusters leads to the rotation of 13.9° of the CDW vector with respect to the crystal lattice[14] (Fig. 2c), which breaks the mirror symmetry and gives rise to the emergence of in-plane chirality. As the rotation can occur clockwise or counter-clockwise, two enantiomers[42,45] with mirrored chiral band structures are formed (Fig. 2) with left and right handedness (Figs. 2c and 2d), thus different geometrical chirality (Supplementary Fig. S3). The two enantiomers (left and right handed) and their rotation angles of $\pm 13.9°$ are directly confirmed by scanning tunneling microscopy (via topography and the corresponding Fourier analysis in Figs. 2c and 2d) and ARPES studies as reconstructed band structures match its own reconstructed Brillouin zone (Supplementary Figs. S15 and S17).

Although circular dichroism has been observed in compounds with chiral crystal and electronic structure[46], it is not observed in the C-CDW phase of $1T$-TaS$_2$, due to the preserved inversion symmetry. Instead, one significant new property of the latter is the chirality-dependent Raman response to circularly polarized light. By definition, a two-dimensional (2D) chiral object is not superimposable on its mirror image by in-plane operations unless it is flipped in three-dimensional space. A 2D chiral crystal necessitates nonzero off-diagonal components for its Raman tensor $\mathbf{R} = \begin{pmatrix} a & c \\ c & b \end{pmatrix}$, because the mirror operation on $\mathbf{R}$ yields $\mathbf{R}' = \begin{pmatrix} a & -c \\ -c & b \end{pmatrix}$, and nonzero $c$ guarantees $\mathbf{R}' \neq \mathbf{R}$, that is, two chiral counterparts. When the tensor elements are complex, the Raman intensities exhibit a contrast in the circular contrarotating configurations, $I_{\sigma^+\sigma^-} \neq I_{\sigma^-\sigma^+}$, where $\sigma^i\sigma^s$ ($i, s = +, -$) represents the helicities for the incident and scattered photons (see the measurement schematic in Fig. 3a). More importantly, the two types of chiral structures are distinguishable by the reversed contrast, because $I_{\sigma^+\sigma^-} = I'_{\sigma^-\sigma^+}$ and $I_{\sigma^-\sigma^+} = I'_{\sigma^+\sigma^-}$, where $I$ and $I'$ represent the Raman intensities measured for opposite chiral structures. More detailed discussions can be found in Supplementary Note S5.

This type of Raman response is realized in 1$T$-TaS$_2$ in the C-CDW phase, because its point group $C_{3i}$ features $E_g$ phonon modes with Raman tensors in the above form[47-49], and its absorbing nature at the excitation photon energy[50] makes the Raman tensors complex. Its Raman spectra show a large number of $A_g$ and $E_g$ phonon modes associated with the CDW superlattice[47-49] (Fig. 3b), consistent with our *ab initio* calculations on the star-of-David structure (Supplementary Fig. S4). The $A_g$ modes are only detected in the two equivalent circular corotating configurations ($I_{\sigma^+\sigma^+} = I_{\sigma^-\sigma^-}$), whereas the $E_g$ modes exhibit $I_{\sigma^+\sigma^-} \neq I_{\sigma^-\sigma^+}$ as expected. Only the $E_g$ modes can reflect the chirality because they correspond to two degenerate vibrations with clockwise or counterclockwise circling motions of atoms, i.e., two chiral phonon modes with opposite angular momentum (Supplementary Fig. S5).

Focusing on the $E_g$ modes in the C-CDW phase, the contrasting Raman response $I_{\sigma^+\sigma^-} \neq I_{\sigma^-\sigma^+}$ was reproduced in multiple samples, divided into two groups with opposite chirality (Supplementary Fig. S6). Each of them shows homogenous properties. Fig. 3c shows the Raman spectra of Samples A and B with opposite chirality at 4 K, which are almost identical for exchanged $\sigma^i$ and $\sigma^s$. They are indistinguishable if measured in the linear or circular corotating polarization configurations (Supplementary Fig. S7). As a more direct test of the chirality, the spectra taken on the two faces of Sample C agree perfectly when $\sigma^i$ and $\sigma^s$ are swapped (see Fig. 3d and more details in Supplementary Fig. S8). These results establish a correspondence between the two types of photon helicity-dependent Raman scattering and distinct crystal structures with opposite chirality.

We further investigate the CDW origin of the chirality in the electronic structure and photon-helicity-dependent Raman signal by temperature-dependent ARPES and Raman scattering (Fig. 4). When raising the sample temperature to 370 K (in I-CDW) and across the NC-CDW to I-CDW transition, the observed band structure becomes quite simple (Fig. 4a, Supplementary Figs. S16 and S24) compared to that of the C-CDW state measured at 21 K (Fig. 4b). Specifically, the observed CECs do not show chiral structure anymore and mirror symmetry is restored (one mirror plane is marked by dotted magenta line in Fig. 4a(ii)). This indicates a chiral-to-achiral transition since the rotation angle between the CDW lattice and crystal lattice relaxes to 0° and the mirror symmetry restores via changing temperature, accompanying the disappearing with the geometrical chirality (Supplementary Fig. S3). We note that chiral electronic structure also exists in the NC-CDW phase (between C-CDW and I-CDW) featuring a rotation of ~ 12°[14], though the windmill shape is not as evident as that of the C-CDW (Supplementary Figs. S16 and S27). The disappearing of the chirality also manifests in the temperature-dependent Raman scattering data. Fig. 4d shows that when temperature is raised from 349 K to 355 K across the transition from the NC-CDW to the I-CDW phase, the contrast between the $\sigma^+\sigma^-$ and $\sigma^-\sigma^+$ spectra disappears. This is because the I-CDW phase features $E_g$ Raman tensors with only diagonal or off-diagonal components, but not both (see Supplementary Note S5). We plot the temperature dependence of the integrated intensity of the normalized differential spectra $(I_{\sigma^+\sigma^-}-I_{\sigma^-\sigma^+})/(I_{\sigma^+\sigma^-}+I_{\sigma^-\sigma^+})$ over 20-300 cm$^{-1}$ (Fig. 4d(iii)) which clearly demonstrates the correlation between the chiral Raman response and the chiral CDW phase. The reversibility exhibited in the warming and cooling processes demonstrates efficient control of the electronic chirality as well as Raman responses by temperature. The simultaneous chiral transition in the electronic structure and Raman response thus proves their common lattice origin as the chiral CDW order.

Our observation of chiral Raman spectra points to a new type of optical response from chiral crystal structures. We derive a complete list of point groups with this form of Raman tensor (Supplementary Table S1), providing a guide to search for 2D chiral crystals. The uniqueness of the chirality in 1$T$-TaS$_2$ lies in its CDW origin, which makes it susceptible to a variety of tuning parameters, such as temperature[12-14], pressure[20], and doping[21-23]. We note that Raman optical activity, which measures the difference in the Raman intensity for either incident or scattered $\sigma^+$ and $\sigma^-$ polarized light from chiral molecules, peptides and proteins (chiral structures in 3D), is several orders of magnitude weaker[51].

In summary, we have discovered chiral electronic structure for both enantiomers in 1$T$-TaS$_2$ in the C-CDW phase. We further show that the chirality naturally originates from in-plane mirror symmetry breaking associated with the commensurate lock-in of the star-of-David clusters. The emergent chiral crystal structure leads to striking photon-helicity-dependent Raman peaks at different chirality. With elevating temperature to the I-CDW phase, mirror symmetry is restored, realizing a chiral-to-achiral transition. Future studies can be carried out on the exotic physics of domain walls between different enantiomer domains, manipulation of the chirality[42,52], and the intricate interplay between chiral order and other rich physics (CDW, Mott insulator, QSL, superconductivity, etc; Fig. 1c). The anomalous Raman response of the C-CDW phase hints other fascinating optical properties may exist in compounds with chiral CDW, which opens up the new revenue for new physics and device applications.

**References:**
1 Wagnière, G. H. On Chirality and the Universal Asymmetry Reflections on Image and Mirror Image (Wiley-VCH, 2007).
2 W.J., L. & I.W., W. Chirality in Natural and Applied Science (2002).

*Methods*

**Single crystal growth**: High-quality single crystals of 1*T*-TaS$_2$ were grown by the chemical vapor transport method with iodine as a transport agent. The high-purity Ta (3.5 N) and S (3.5 N) were mixed in chemical stoichiometry and heated at 850 ℃ for 4 days in an evacuated quartz tube. The harvested TaS$_2$ powders and iodine (density: 5 mg/cm$^3$) were then sealed in another quartz tube and heated for 2 weeks in a two-zone furnace, in which the source zone and growth zone were fixed at 900 ℃ and 800 ℃, respectively. The tubes were rapidly quenched in cold water to ensure retaining of the 1*T* phase.

**Angle-resolved photoemission spectroscopy**: ARPES measurements were conducted at BL03U and 'Dreamline' of Shanghai Synchrotron Radiation Facility, P. R. China, and BL7 of Advanced Light Source, USA. Single-crystals were *in-situ* cleaved at low temperature to expose clean surface for measurements. The overall energy and angular resolutions are 10-20 meV and 0.2°, respectively. The photon-energy, polarization, and temperature-dependent measurements were performed. Preliminary ARPES studies were carried out at BL13U of National Synchrotron Radiation Laboratory, and lab-based helium-lamp ARPES system in ShanghaiTech University.

**Scanning tunneling microscopy**: STM experiments were carried out with a Unisoku low-temperature STM at the base temperature of 4.3 K. 1*T*-TaS$_2$ single crystal samples were cleaved at ~ 77 K, and then transferred into the STM head for measurements.

**Raman scattering measurements:** Polarized Raman scattering microspectroscopy was performed in the backscattering geometry. A schematic for the setup is shown in Supplementary Fig. S9. A ×40 objective was used both for focusing the incident beam onto the sample to a spot size of ~1 $\mu$m and for collecting the scattered light. By setting the fast axis of the quarter-wave plate to $\pm 45°$ with respect to the first polarizer and setting the second polarizer perpendicular or parallel to the first polarizer, the circularly corotating ($\sigma^+\sigma^+$ and $\sigma^-\sigma^-$) or contrarotating ($\sigma^+\sigma^-$ and $\sigma^-\sigma^+$) configuration was achieved, respectively. Measurements in the collinear (XX) or cross (XY) polarization configurations were performed by setting the fast axis of the quarter-wave plate parallel to the first polarizer. A pair of notch filters efficiently suppressed the Rayleigh scattering and enabled low-wavenumber measurements down to 15 cm$^{-1}$. The signal was detected using a Princeton Instruments grating spectrometer equipped with a liquid-nitrogen-cooled charge-coupled device. The samples were mounted in a Montana Instruments Cryostation for temperature control from 4-360 K. An excitation laser wavelength of 532 nm was used. The incident power was limited to be $\leqslant 0.5$ mW. All samples used in the Raman measurements were bulk flakes thicker than 50 nm, confirmed by atomic force microscopy.

***ab-initio* calculations**: We calculated the band structure of the 1$T$-TaS$_2$ monolayer in the star of David charge density wave phase by the density-functional theory. The star-of-David structure exhibits inversion and three-fold rotational symmetry ($C_3$) that belongs to the space group No. 147. The structure information in our calculation is given in the following. We performed DFT calculation in the framework of the generalized gradient approximation[53]). The phonon calculation performed by phonopy package[54] with Vienna *ab-intio* package[55]. Spin-orbit coupling was included in all calculations. The structure unfolding calculation performed by VASPKIT[56].

**Acknowledgements**

We wish to thank Dr. W. J. Shi, Dr. Y.Y.Y. Xia, and Prof. G. Li for insightful discussions on tight-binding model with the Slater-Koster method; and Dr. J. Xu and Prof. L.B. Gao for atomic force microscopy measurements on the samples used in the Raman study; and Dr. S. Wu and Prof. S. W. Wu for optical reflectivity measurements. We acknowledge the following ARPES beamlines: BL03U and 'Dreamline' of Shanghai Synchrotron Radiation Facility, and BL13U of National Synchrotron Radiation Laboratory, and BL7 of Advanced Light Source. We also acknowledge the Analytical Instrumentation Center of ShanghaiTech University for X-ray and Laue diffraction measurements. This work was supported by the Shanghai Municipal Science and Technology Major Project (grant No. 2018SHZDZX02 to Y.L.C. and Z.K.L.), the National Natural Science Foundation of China (grant Nos. 11634009 and 11674229 to Y.L.C. and Z.K.L., grant Nos. 11674326 and 11874357 to J.J.G, X.L. and Y.P.S., grant No. 11774151 to X.X., grant No. 12004248 to H.F.Y., and A3 Foresight Program 51861145201), the National Key R&D Program of China (grant No. 2017YFA0305400 to Z.K.L., grant Nos. 2018YFA0307000 and 2017YFA0303201 to X.X., and grant No. 2016YFA0300404 to X.L. and Y.P.S.), the Joint Funds of the National Natural Science Foundation of China and the Chinese Academy of Sciences Large-Scale Scientific Facility (grant Nos. U1832141, U1932217 and U2032215 to X.L. and Y.P.S.), the Key Research Program of Frontier Sciences (grant No. QYZDB-SSW-SLH015 to J.J.G, X.L. and Y.P.S.), Excellence and Scientific Research Grant of Hefei Science Center of CAS (grant No. 2018HSC-UE011 to J.J.G, X.L. and Y.P.S.), Shanghai Sailing Program (grant No. 20YF1430500 to H.F.Y.). B.Y. acknowledges the financial support by the European Research Council (ERC Consolidator Grant ``NonlinearTopo'', No. 815869) and the MINERVA Stiftung with the funds from the BMBF of the Federal Republic of Germany.

**Author contributions**

Z.K.L., X.X., Y.L.C., and B.H.Y. conceived the research. H.F.Y. performed the ARPES experiments with the assistance of C.C., A.J.L., K.H., M.X.W., and L.X.Y., K.Y.H. performed the Raman experiments with the assistance of G.L., J.K. performed the ab-initio calculations with the assistance of Y.Z.L., S.W.S and S.C.Y. performed the STM measurements, J.J.G., X.L., and Y.P.S. synthesized the 1T-TaS2 crystals. S.H.Z. and J.P.L. conducted preliminary calculations.

**Competing interests**

The authors declare no competing interests.

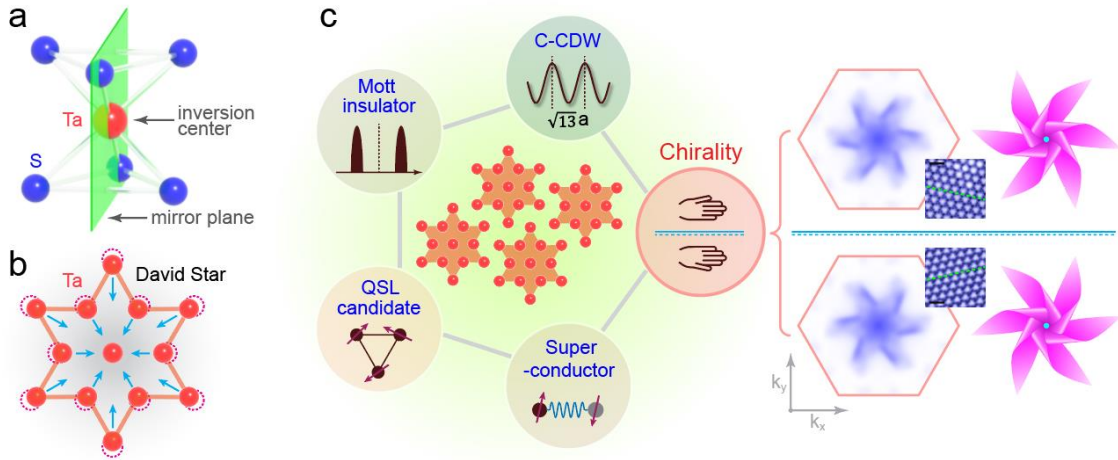

**Fig. 1. a**. Schematic of crystal structure of 1$T$-TaS$_2$ in the normal state, which is achiral as both inversion and mirror symmetries are preserved. **b**. Sketch of the David-Star clusters in the C-CDW state. **c**. Schematic of interesting physical properties of 1$T$-TaS$_2$ and the hidden chiral order, which is featured by the chiral CDW and windmill-shape band structure with broken mirror symmetry.

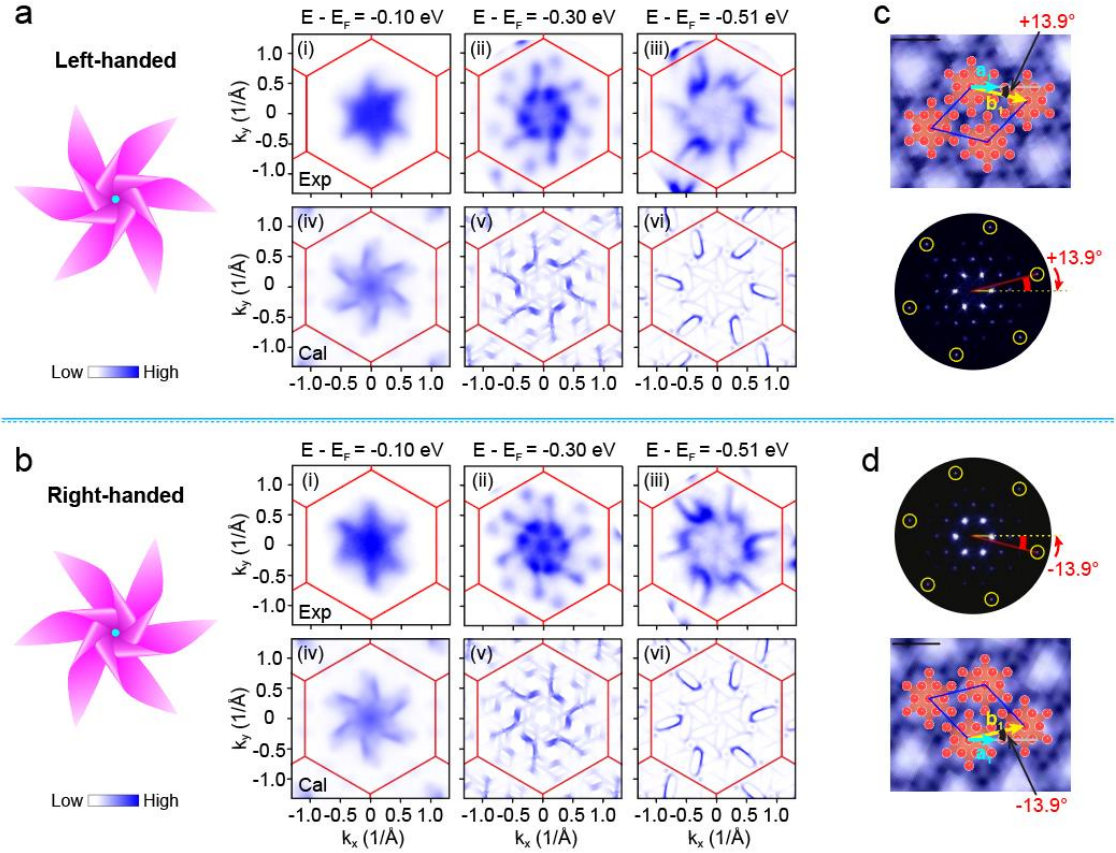

**Fig. 2. a.** Windmill-shape band structures of the chiral-left domain of 1*T*-TaS$_2$ in the C-CDW phase. (i)-(iii) Representative CECs at -0.1 eV, -0.3 eV, and -0.51 eV measured with 94 eV linearly-polarized photons. (iv)-(vi) Calculated CECs at the corresponding energies. **b.** Same as **a** but measured on the chiral-right domain. **c.** Topography (upper) and its Fourier transform results (lower) on the chiral-left domain via STM (sample bias: 500 mV, tunnelling current: 19 pA, sample temperature: 4.3 K). Scale bar 1 nm. The unit vector of the C-CDW (yellow arrow) rotates clockwise by 13.9° with respect to the lattice vector (cyan arrow). **d.** Same as **c** but measured on the chiral-right domain. The unit vector of the C-CDW (yellow arrow) rotates counter-clockwise by 13.9° with respect to the lattice vector (cyan arrow).

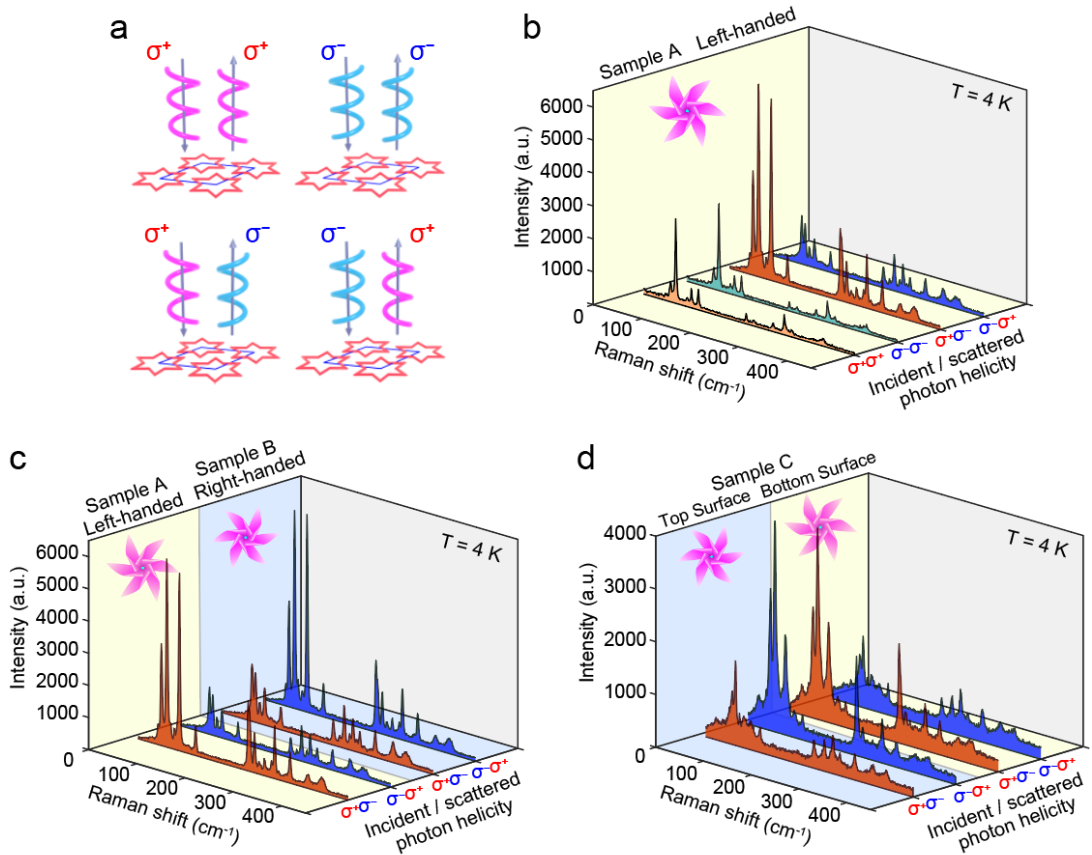

**Fig. 3.** Photon-helicity-dependent Raman response in C-CDW phases of 1$T$-TaS$_2$. **a**. Schematic of polarization-resolved Raman response measurements. **b**. Raman spectra of Sample A measured in the circular corotating ($\sigma^+\sigma^+$ and $\sigma^-\sigma^-$) and circular contrarotating ($\sigma^+\sigma^-$ and $\sigma^-\sigma^+$) polarization configurations. **c**. Raman spectra of Sample A and B showing opposite $\sigma^+\sigma^-/\sigma^-\sigma^+$ response. **d**. Swapped $\sigma^+\sigma^-/\sigma^-\sigma^+$ Raman response on the opposite sides of Sample C. All Raman spectra taken at T = 4 K.

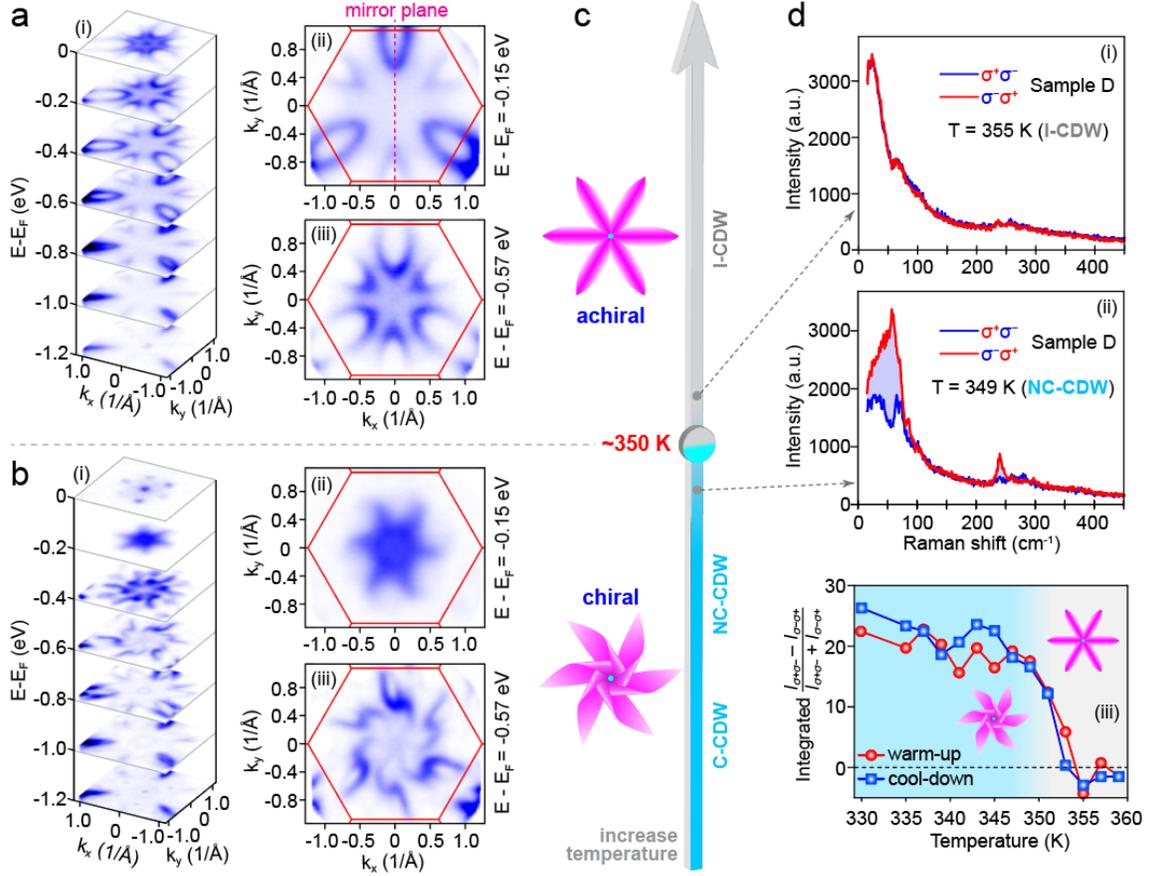

**Fig. 4.** Chiral-to-achiral transition when elevating temperature to cross ~ 350 K. **a-b**. Comparison of band structures of 1$T$-TaS$_2$ at 21 K (C-CDW) (**a**) and 370 K (I-CDW) (**b**). **c**. Schematic illustration of chiral-to-achiral transition at ~ 350 K. **d**. Circular contrarotating Raman response of Sample D at 349 K (i) and 355 K (ii). The shaded area in (ii) highlights $I_{\sigma^+\sigma^-} - I_{\sigma^-\sigma^+}$ at 349 K. (iii) Temperature dependence of the normalized differential spectra $(I_{\sigma^+\sigma^-} - I_{\sigma^-\sigma^+})/(I_{\sigma^+\sigma^-} + I_{\sigma^-\sigma^+})$ integrated over 20-300 cm$^{-1}$, clearly showing the chiral-to-achiral transition occurs ~ 350 K.